\documentclass[sigconf]{acmart}
\usepackage{soul}
\usepackage{xcolor}
\usepackage{booktabs}
\sethlcolor{yellow}
\usepackage{enumitem}
\usepackage{graphicx}
\usepackage{amsmath}
\usepackage{pgfplotstable}
\usepackage{listings}

\AtBeginDocument{%
  }

\renewcommand\footnotetextcopyrightpermission[1]{} 
\setcopyright{none} 
\acmVolume{}
\acmNumber{}
\acmArticle{}

\definecolor{codebg}{RGB}{245,245,245}  

\begin{document}

\title{Bridging the Urban Divide: Adaptive Cross-City Learning for Disaster Sentiment Understanding}


\author{Zihui Ma}
\authornote{Co-leading authors.}
\affiliation{%
 \institution{New York University}
 \city{New York}
 \country{USA}}
\email{zihuima@nyu.edu}

\author{Yiheng Chen}
\authornotemark[1]
\affiliation{%
 \institution{University of Alabama}
 \city{Tuscaloosa}
 \country{USA}}
\email{ychen226@crimson.ua.edu}

\author{Runlong Yu}
\affiliation{%
 \institution{University of Alabama}
 \city{Tuscaloosa}
 \country{USA}}
\email{ryu5@ua.edu}

\author{Afra Izzati Kamili}
\affiliation{%
 \institution{New York University}
 \city{New York}
 \country{USA}}
\email{ak12151@nyu.edu}

\author{Fangqi Chen}
\affiliation{%
 \institution{New York University}
 \city{New York}
 \country{USA}}
\email{fc2567@nyu.edu}

\author{Zhaoxi Zhang}
\affiliation{%
 \institution{University of Florida}
 \city{Gainesville}
 \country{USA}}
\email{zhang.zhaoxi@ufl.edu}

\author{Juan Li}
\affiliation{%
 \institution{Google LLC}
 \country{USA}}
\email{leahlijuan1@gmail.com}

\author{Yuki Miura}
\authornote{Corresponding author.}
\affiliation{%
 \institution{New York University}
 \city{New York}
 \country{USA}}
\email{yuki.miura@nyu.edu}


\renewcommand{\shortauthors}{Zihui Ma, Yiheng Chen, et al.}

\begin{abstract}
Social media platforms provide a real-time lens into public sentiment during natural disasters; however, models built solely on textual data often reinforce urban-centric biases and overlook underrepresented communities. This paper introduces an adaptive cross-city learning framework that enhances disaster sentiment understanding by integrating mobility-informed behavioral signals and city similarity-based data augmentation. Focusing on the January 2025 Southern California wildfires, our model achieves state-of-the-art performance and reveals geographically diverse sentiment patterns, particularly in areas experiencing overlapping fire exposure or delayed emergency responses. 
We further identify positive correlations between emotional expressions and real-world mobility shifts, underscoring the value of combining behavioral and textual features. Through extensive experiments, we demonstrate that multimodal fusion and city-aware training significantly improve both accuracy and fairness. Collectively, these findings highlight the importance of context-sensitive sentiment modeling and provide actionable insights toward developing more inclusive and equitable disaster response systems.
\end{abstract}

\maketitle

\section{Introduction}

Social inequities often intensify during disasters, shaping how different communities access, interpret, and share information under crisis conditions \cite{chen2022investigation,mansourihanis2025equity}. 
Online social platforms have become critical tools for monitoring public emotion and awareness, yet their data often reflect a biased sample of society\cite{fan2020spatial, samuels2020deepening}. Urban and higher-income regions tend to dominate online discourse, while rural or disadvantaged communities are underrepresented \cite{zhang2021revealing, ma2024surveying, ma2025analyzing, ramakrishnan2022examining}. This imbalance distorts assessments of collective sentiment and resilience, thereby limiting decision-makers’ ability to understand the actual needs of the community during disaster recovery. Addressing this digital divide is essential for building fair, inclusive, and context-aware disaster intelligence systems~\cite{luo2025geo,yu2025environmental}.

Recent advances in natural language processing have substantially improved the ability to analyze social media text. Transformer-based models such as BERT \cite{vaswani2017attention, devlin2019bert} and RoBERTa \cite{liu2019roberta} can capture contextual and semantic nuances, enabling fine-grained sentiment and topic classification. These capabilities support a more comprehensive understanding of disaster-related communication, public awareness, and emotional dynamics \cite{liu2021crisisbert, paul2023fine, egger2022topic, grootendorst2022bertopic}. However, language models inherit biases from socially and geographically imbalanced training data, often amplifying existing inequities\cite{zheng2024reducing, weidinger2021ethical, navigli2023biases,gallegos2024bias}. Large language models (LLMs) further extend reasoning capacity but introduce additional limitations, including high computational cost, sensitivity to prompting, and reliance on commercial APIs \cite{zhao2023survey, yang2025comprehensive,li2025llms,yu2025survey}. In addition, their slow inference, limited transparency, and privacy risks can exacerbate uneven performance across regions and populations \cite{dos2024identifying, choi2025advantages,chen2025empirical}. 
These limitations raise a central question: can we design adaptive models that integrate behavioral evidence to mitigate \emph{online representation bias} and improve sentiment robustness across unevenly represented cities during disasters?

To address this challenge, we propose an \textbf{adaptive cross-city learning framework} that aligns online sentiment with real-world behavior. The framework operates at two levels. The first layer models inter-city similarity using socioeconomic and demographic attributes to perform data augmentation for underrepresented regions. This process transfers knowledge from data-rich to data-sparse areas, improving generalization and fairness~\cite{luo2025great}.
Rather than assuming behavioral equivalence across cities, inter-city similarity is used as a weak structural prior to stabilize learning in low-resource settings~\cite{yu2025physics}, while preserving city-specific sentiment signals through similarity-weighted adaptation. The second layer integrates human mobility features to represent behavioral responses such as evacuation and return. These features provide evidence to textual sentiment, revealing how people act in the physical world when faced with disasters. By combining linguistic and behavioral representations, the model establishes a dynamic link between \textit{words and actions}~\cite{li2025llms}, reducing representation bias in language model predictions and yielding more stable and geographically calibrated estimates of collective sentiment. Our key contributions are threefold:
\begin{itemize}[nosep]
\item \textbf{Adaptive cross-city learning framework.} We introduce a two-layer learning architecture that combines cross-city similarity modeling and multimodal (text--mobility) feature fusion to mitigate bias in language-model-based sentiment classification.
\item \textbf{Dynamic word-to-action analysis.} We investigate how public emotions expressed online correlate with real-world behavioral patterns, as captured through mobility signals, during wildfire events.
\item \textbf{Fairness and representation improvement.} We demonstrate through empirical evaluation that integrating mobility embeddings enhances both prediction accuracy and calibration, especially for underrepresented cities.
\end{itemize}

\section{Related Work}
\subsection{\mbox{Social inequality in disaster communication}}
Disaster communication is an interactive and iterative process of exchanging information among stakeholders throughout the phases of mitigation, preparedness, response, and recovery \cite{reynolds2005crisis,haddow2013disaster,ma2024surveying}. Effective and accurate communication enhances public preparedness and reduces disaster risks, serving as a cornerstone of successful emergency management \cite{bradley2016effectiveness,savoia2013communications}. In recent years, social media platforms such as X (formerly Twitter), Facebook, and Instagram have become key venues for individuals to share their preparedness, perceptions, and experiences of disasters, accelerating the online circulation of crisis information. These crowd-sourced data have been increasingly leveraged to analyze public responses \cite{dong2021social,ma2024investigating,li2020leveraging}, including rescue coordination \cite{li2023environmental}, sentiment expression \cite{karimiziarani2025natural,gupta2025sentimentmapper,babu2025advanced}, and disaster information dissemination \cite{fan2025dynamics,ma2024investigating,al2025information}.

However, concerns about representativeness persist, as access to and engagement with digital platforms remain uneven across social and demographic groups. Social media users constitute a self-selected population rather than a random sample of society \cite{cesare2019understanding,chen2022investigation,dargin2021vulnerable}. Prior studies highlight that marginalized or digitally disconnected populations are systematically underrepresented in online disaster discourse \cite{ma2024surveying,fan2020spatial,wang2025mining}. For example, \citet{chen2022investigation} found that White users are overrepresented in disaster-related Twitter discussions, whereas Hispanic and Black populations are significantly underrepresented, with disparities varying by region. Similarly, \citet{mavrodieva2021social} observed that elderly populations face significant barriers to digital participation. Such imbalances, driven by social and technological inequality, can distort how public perception and sentiment are portrayed during crises. Therefore, mitigating these biases is essential for improving the reliability and inclusiveness of social media–based disaster research.

\vspace{-0.2cm}
\subsection{Fairness-aware sentiment analysis}

Sentiment analysis has long been a cornerstone of understanding public discourse, evolving from lexicon and feature-based methods to deep contextual encoders and, more recently, LLMs \cite{Wankhade2022SentimentSurvey,zhao2023survey}. Transformer-based models such as BERT\cite{devlin2019bert} and RoBERTa \cite{liu2019roberta} capture nuanced linguistic cues and have become standard for short, informal social media posts, with domain-specific variants like BERTweet achieving state-of-the-art accuracy on Twitter data and crisis-related text classification \cite{Nguyen2020BERTweet,liu2021crisisbert,paul2023fine}. However, despite impressive overall performance, these models often struggle to generalize across domains, regions, or events—precisely the settings where crisis communication unfolds. Benchmarks such as CrisisBench \cite{Alam2021CrisisBench} and WILDS \cite{Koh2021WILDS} reveal that sentiment classifiers trained in one geography or event type degrade sharply when applied elsewhere. This instability highlights a fundamental challenge: sentiment models capture what people say but not where or under what circumstances they speak.

Fairness-aware learning offers a complementary perspective, asking not only whether predictions are accurate on average, but also whether they are equitable across subpopulations \cite{Mehrabi2021BiasFairnessSurvey,Bansal2022NLPBiasFairness}. In natural language processing, bias can manifest through lexical stereotypes and stereotyped associations in word or sentence representations \cite{sun2019mitigating,Czarnowska2021QuantifyingSocialBiasesNLP}, unbalanced error rates and fairness metrics across demographic groups \cite{Mehrabi2021BiasFairnessSurvey,Czarnowska2021QuantifyingSocialBiasesNLP}, or the underrepresentation of certain communities in training data \cite{cesare2019understanding,chen2022investigation,ramakrishnan2022examining}. As models scale to LLMs, such biases are amplified by large and opaque pretraining corpora \cite{bender2021dangers,weidinger2021ethical,gallegos2024bias,navigli2023biases,Nadeem2021StereoSet}. This has motivated a rich body of fairness-aware methods at three levels: (i) data-level augmentation that edits sensitive attributes or balances samples \cite{Kaushik2020CDA}; (ii) representation-level debiasing through projection- or adversarial-based removal of protected information from contextual embeddings \cite{Ravfogel2020INLP,Kaneko2021DebiasingContextualized}; and (iii) objective-level optimization to improve worst-group generalization via distributionally robust learning \cite{Sagawa2020DRO}. While effective in controlled settings, these approaches often assume static domains and overlook spatial or behavioral disparities that shape how different groups express emotion online.

Recent studies in crisis informatics have begun to expose these contextual gaps. Social media users in urban, higher-income areas are overrepresented during disasters, whereas rural and vulnerable communities are often digitally invisible \cite{fan2020spatial,samuels2020deepening,zhang2021revealing,mavrodieva2021social,dargin2021vulnerable}. Traditional sentiment models (trained purely on textual features) thus risk amplifying structural inequalities by misrepresenting the collective emotional state of affected populations. Bridging this gap requires not only linguistic understanding but also contextual grounding: integrating external evidence such as mobility, demographics, and hazard exposure to interpret online emotion through the lens of real-world behavior.

\begin{figure*}[htbp]
    \centering
    \includegraphics[width=0.9\textwidth]{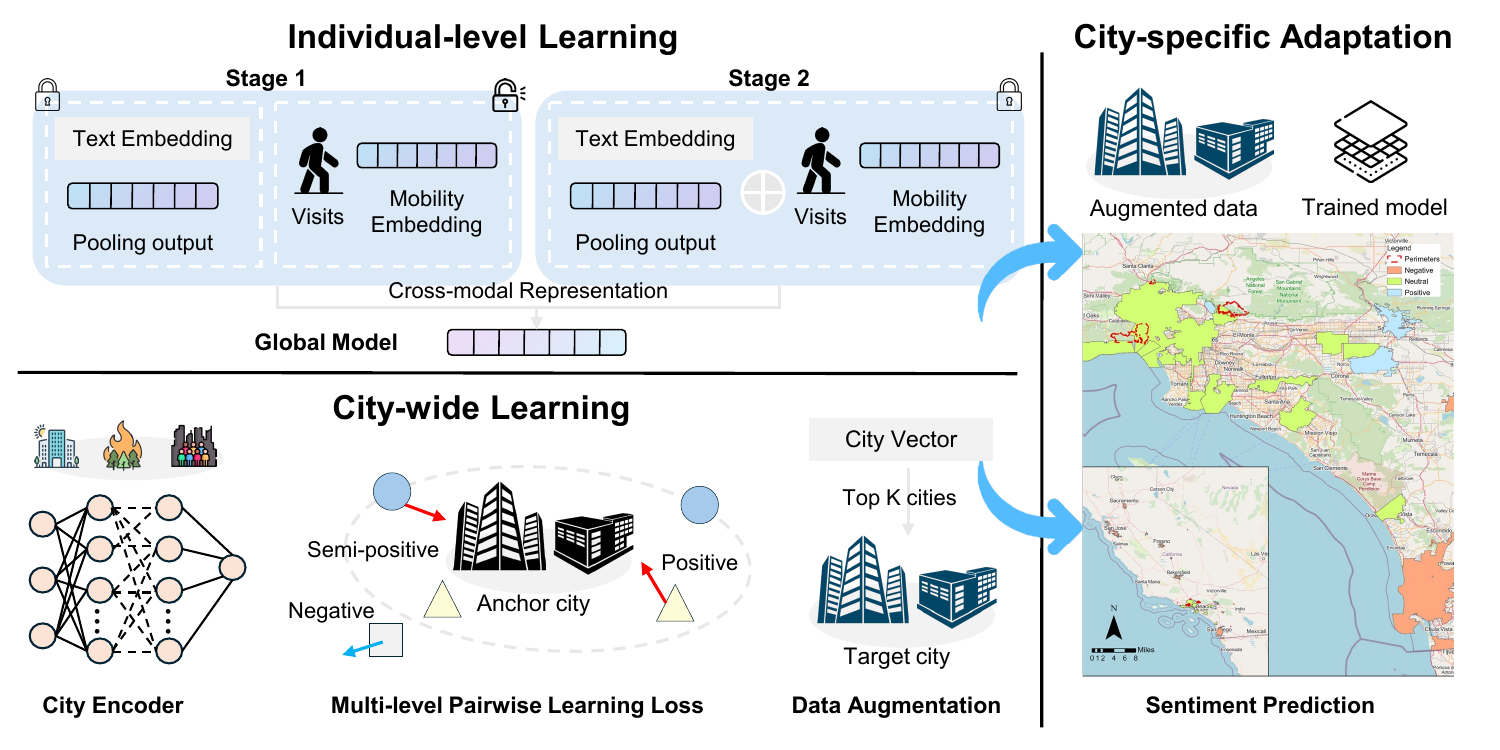}
    \vspace{-5pt}
    \caption{The overall framework.}
    \vspace{-10pt}
    \label{fig: framework}
\end{figure*}

\section{Data Preparation}
\label{sec:data}
We construct two main categories of data to train our framework: (1) Tweets and Mobility Multimodal Data, which provide tweet-level inputs for the Individual-level Learning Layer, and (2) City Static Data with Risk Level, which provide city-level inputs for the City-wide Learning Layer.

\subsection{Tweets and mobility multimodal data}

This component integrates two complementary data sources that jointly capture linguistic and behavioral signals. 
Tweets provide text-based observations of public sentiment, while mobility traces capture real-world human movement patterns. 
\vspace{-0.25cm}
\paragraph{Tweets data.}
Wildfire-related tweets were collected via the Brandwatch API \cite{brandwatch2025suite}using keyword-based queries focused on California. 
Tweets with geotags were prioritized for location accuracy; when absent, tweet locations were inferred from textual cues using a large language model~\cite{chen2025empowering, yu2025rag}, and, when necessary, supplemented by user registration metadata.
\vspace{-0.25cm}
\paragraph{Mobility data.}
Anonymized, device-level GPS traces provided by Spectus \cite{spectus2025mobility} were used to describe population movement. 
Raw trajectories were aggregated into daily origin–destination (OD) matrices at the Census tract level, and the resulting travel flows were summarized into city-level mobility features.

We align tweets and mobility at a common city-level unit by mapping Census tracts to Place IDs via the National Historical Geographic Information System (NHGIS)\cite{ipums_nhgis_v20} Block Group-to-Places crosswalk and matching Census-defined city names with language-model–inferred locations, producing a unified multimodal dataset.

\subsection{City static data with risk level}

This component characterizes the structural and environmental attributes of each city, providing essential inputs for the City-wide Learning Layer. 
Socioeconomic indicators capture long-term demographic and infrastructural differences, while disaster risk scores quantify each city's exposure to wildfire hazards.

\vspace{-0.25cm}
\paragraph{Socioeconomic data.}
Tract-level static features were obtained from the American Community Survey (ACS). 
Highly collinear variables were removed through variance inflation factor (VIF) analysis, and the remaining attributes were combined with urbanization rates derived from population datasets to construct tract-level socio-economic feature vectors.

\vspace{-0.25cm}
\paragraph{Risk data.}
Wildfire-related hazard information was extracted from the FEMA National Risk Index \cite{fema_nri}, which provides hazard-specific risk scores at the Census tract level across the United States. 
Only the wildfire risk category was retained to align with the study’s focus on wildfire-related sentiment and exposure.

Both datasets were aggregated to the city level using the same tract-to-city mapping procedure described above, with population-weighted averaging applied to derive city-level static and risk feature vectors for training the City-wide Learning Layer.

\section{Methods}

\subsection{Framework overview}

Our objective is to build \emph{city-specific} sentiment models that remain globally reliable while adapting to the unique social, mobility, and hazard contexts of individual cities. Models trained solely at the global level tend to overfit data-rich urban centers and yield unstable or poorly calibrated predictions elsewhere~\cite{sun2025domain}. We propose an adaptive cross-city framework that uses inter-city similarity as a weak structural prior and grounds sentiment inference in real-world mobility evidence (Figure~\ref{fig: framework}).

At the core of the framework is a city-specific adaptation mechanism. For each target city, we start from a globally trained multimodal model and refine it using an augmented dataset that combines local tweets with similarity-weighted samples from structurally related cities, where similarity is learned from socioeconomic and wildfire-risk attributes. In parallel, tweet-level mobility features provide behavioral context, linking what people \emph{say} online with how they \emph{act} during wildfire disruptions, yielding locally calibrated predictors even in data-scarce cities. The framework consists of two complementary layers:

\textbf{(1) Individual-level learning layer (ILL). }  
This layer operates on multimodal tweet data. It fuses textual semantics with city-level mobility features to train a geographically aware global model. By aligning linguistic expressions with mobility-driven behavioral patterns, the model learns transferable cross-modal representations that capture how similar expressions may signal different emotional states under different city contexts.

\textbf{(2) City-wide learning layer (CLL).}  
This layer operates on city-level socioeconomic and hazard attributes, learning a city-level encoder in which similarity reflects shared structural conditions. While it does not explicitly model local media or governance factors, the representation is intended to mitigate representation bias from uneven data availability. The resulting embeddings guide a similarity-weighted augmentation process that improves data efficiency while preserving city-specific sentiment patterns.

Together, these layers form a unified architecture for city-specific adaptation that balances global robustness with local sensitivity, improving sentiment stability across unevenly represented cities.

\vspace{-0.5cm}
\subsection{Individual-level learning layer}
\label{sec:individual-layer}


At the individual level, textual semantics and city mobility are jointly modeled to enable context-aware sentiment prediction. We adopt a Transformer-based encoder-only architecture as the text encoder, which yields bidirectional contextual representations and fixed-length embeddings in a single forward pass—more efficient and suitable for classification than autoregressive decoding. It is also noted that we intentionally employ a lightweight Transformer model rather than a large language model to better align model complexity with the characteristics of our dataset, where tweets are typically short (on average 16 tokens, with the longest sequence containing fewer than 40 tokens). This compact architecture is sufficient enough to capture the available semantic variability while substantially reducing memory footprint and training/inference time. Moreover, it supports scalable deployment in resource-constrained settings (e.g., city-level or agency systems) without compromising predictive performance. Each tokenized tweet is then mapped to a sequence of contextualized hidden states and aggregated via a pooling operation into a fixed-size tweet embedding. In parallel, the corresponding mobility feature vector is processed through a small feed-forward network, referred to as the mobility encoder, to obtain a dense mobility embedding. The embedding dimension is chosen to be comparable to that of the text representation, so that the mobility signal is not numerically dominated during fusion.

Formally, let $x^{(t)}$ and $x^{(m)}$ denote the textual and mobility inputs, respectively. A text encoder maps $x^{(t)}$ to contextual token representations $H = E_t(x^{(t)})\in\mathbb{R}^{L\times d}$, from which we derive a fixed-length tweet embedding by applying an affine transformation followed by $\tanh$ to the \texttt{[CLS]} representation:
$
h_t = \tanh\!\big(W_p\,H_{[\texttt{CLS}]} + b_p\big).
$
In parallel, a mobility encoder produces a mobility embedding $h_m = E_m(x^{(m)})$. The two embeddings are concatenated and transformed by a fusion network $F(\cdot)$ into a joint representation 
$
h_f = F([h_t; h_m]).
$
A prediction head $C(\cdot)$ outputs sentiment probabilities $\hat{y} = C(h_f)$, and the model is trained with a standard cross-entropy loss. To construct a geographically aware base model, we apply a two-stage training strategy:

\emph{Weakly supervised pretraining on LLM-generated labels.}  
We first train the multimodal classifier on large-scale, city-agnostic data whose sentiment labels are generated by a large language model. Although noisy, these weak labels are abundant and allow the model to align textual and mobility modalities while learning coarse sentiment patterns. During this stage, the text encoder is frozen to preserve its linguistic knowledge, while the mobility encoder, fusion network, and classifier adapt to the weak supervision.

\emph{Supervised refinement on human-annotated labels.}  
We continue training on a smaller, high-quality dataset labeled by human annotators. These reliable labels refine the decision boundary and mitigate bias introduced in Stage~1. To calibrate the classifier without overwriting the learned multimodal alignment, both encoders are kept frozen and only the fusion network and classifier are updated.

We denote the resulting model after Stage~2 by $f_\theta$ and refer to it as the global model. In the following subsections, we adapt $f_\theta$ to each city using the city-wide similarity structure.

\subsection{City-wide learning layer}

Our CLL focuses on modeling relationships among cities and borrows from recent augmentation-adaptive representation learning frameworks \cite{DBLP:journals/corr/abs-2509-14563} to learn a similarity space over socio-environmental and wildfire-risk attributes, which is then used to guide data augmentation for underrepresented cities.

Formally, each city $c \in \mathcal{C}$ is represented by a static feature vector $f_c \in \mathbb{R}^{D}$, constructed from socioeconomic and wildfire-risk attributes described in Section~\ref{sec:data}. A city encoder $\phi: \mathbb{R}^{D} \rightarrow \mathbb{R}^{d}$ maps these features to a low-dimensional embedding $z_c = \phi(f_c)$. To capture graded similarity among cities, we adopt a \emph{risk-aware triplet-style contrastive objective}. For an anchor city $A$ and three other cities $P, S, N$ (positive, semi-positive, and negative) selected according to wildfire-risk differences, 
the loss encourages risk-consistent relative similarity:
\begin{equation} 
\begin{aligned} 
\mathcal{L} = -\log \sigma\!\bigl( & \lambda [\mathrm{sim}(z_A, z_P) - \mathrm{sim}(z_A, z_S)] \\
&\quad + (1-\lambda)[\mathrm{sim}(z_A, z_S) - \mathrm{sim}(z_A, z_N)] \bigr),
\end{aligned} 
\end{equation}
where $\mathrm{sim}(\cdot,\cdot)$ denotes cosine similarity, $\sigma(u) = 1 / (1 + e^{-u})$ is the logistic sigmoid, and $\lambda \in [0,1]$ balances the two pairwise constraints. Intuitively, the objective enforces
\[
\mathrm{sim}(z_A,z_P) > \mathrm{sim}(z_A,z_S) > \mathrm{sim}(z_A,z_N),
\]
so that cities with very similar risk profiles are closer to the anchor than moderately similar ones, and both are closer than highly dissimilar cities.

For each mini-batch, anchor cities $A$ are uniformly sampled from $\mathcal{C}$. Given the wildfire-risk score $r(c)$, other cities are grouped by their absolute risk difference $|r(c)-r(A)|$ into three sets using thresholds $0 \le \tau_1 < \tau_2$:
\begin{equation}
\begin{aligned}
\mathcal{P}(A) &= \{c \in \mathcal{C} : |r(c)-r(A)| \le \tau_1 \}, \\
\mathcal{S}(A) &= \{c \in \mathcal{C} : \tau_1 < |r(c)-r(A)| \le \tau_2 \}, \\
\mathcal{N}(A) &= \{c \in \mathcal{C} : |r(c)-r(A)| > \tau_2 \}.
\end{aligned}
\end{equation}
If any pool is empty, we fall back to uniform sampling from $\mathcal{C}$ to maintain training stability.

After contrastive pretraining, we perform nearest-neighbor search in the learned embedding space to identify similar cities for each target city $i \in \mathcal{C}$. Let the candidate set be
\begin{equation}
\mathcal{M}(i) = \{\, j \in \mathcal{C} \setminus \{i\} \mid \mathrm{sim}(z_i, z_j) \ge 0 \,\},
\end{equation}
and define the Top-$K$ most similar cities as
\begin{equation}
\mathcal{N}_K(i) = \mathrm{TopK}_{j \in \mathcal{M}(i)}\big(\mathrm{sim}(z_i,z_j)\big).
\end{equation}
We index these embeddings using FAISS for efficient retrieval. The neighbor sets $\mathcal{N}_K(i)$ serve as the basis for similarity-weighted data augmentation during city-specific adaptation, allowing data-sparse cities to borrow information from structurally similar ones rather than from arbitrary large metropolitan areas.

\subsection{City-specific adaptation}

Given the global model $f_\theta$ and the city neighbor sets $\mathcal{N}_K(i)$ obtained from the City-wide Learning Layer, we specialize sentiment prediction to each city via similarity-guided fine-tuning. The key idea is to augment a target city’s labeled tweets with weighted examples from its most similar cities, thereby approximating the target city’s latent distribution while preserving its local characteristics.

For each target city $i$, let $\mathcal{D}_i = \{(x, y)\}$ denote its labeled dataset, where $x$ is the multimodal input and $y \in \{-1, 0, 1\}$ is the sentiment label. To alleviate data sparsity in low-resource cities, we augment $\mathcal{D}_i$ with samples from its Top-$K$ neighbor cities $\mathcal{N}_K(i)$ using similarity-based weights
\begin{align}
\alpha_{ij} &=
\frac{\exp(\gamma\,\mathrm{sim}(z_i,z_j))}
{\sum_{k\in\mathcal{N}_K(i)} \exp(\gamma\,\mathrm{sim}(z_i,z_k))},
\quad j\in\mathcal{N}_K(i),
\label{eq:alpha_weight}
\end{align}
where $\gamma$ controls the sharpness of the similarity distribution. Each sample $(x,y)$ from city $j$ receives weight $w(x,y)=\alpha_{ij}$, while samples from the target city $i$ are assigned weight $w(x,y)=1$.

We partition the parameters of $f_\theta$ as $ \theta=\{\theta_t,\theta_m,\theta_f,\theta_c\},$
corresponding to the text encoder, mobility encoder, fusion network, and classifier, respectively. During city-specific adaptation, we freeze $\theta_t$ and $\theta_m$ to preserve the shared multimodal representations, and fine-tune only $\theta_f$ and $\theta_c$ on the weighted dataset. The adaptation objective minimizes the weighted empirical risk:
\begin{align}
\min_{\theta_f,\theta_c} \mathcal{L}
&= \frac{1}{Z}
\sum_{(x,y)\in\mathcal{D}_i}
\mathcal{L}_{\text{CE}}(f_\theta(x),y) \notag\\
&\quad + \frac{1}{Z}
\sum_{j\in\mathcal{N}_K(i)}
\sum_{(x,y)\in\mathcal{D}_j}
\alpha_{ij}\,\mathcal{L}_{\text{CE}}(f_\theta(x),y),
\label{eq:weighted_loss}
\end{align}
where $\mathcal{L}_{\text{CE}}$ denotes cross-entropy loss and $Z$ is a normalization constant (e.g., the total effective weight).


\section{Experiments and Results}
\subsection{Case background and datasets}

This study centers on Southern California, a region characterized by its large and diverse population, urban–rural heterogeneity, and recurring exposure to wildfire hazards. We focus on the January 2025 wildfires, which collectively led to the evacuation of over 200,000 residents, the destruction of more than 18,000 structures, and the burning of over 57,000 acres \cite{calfire2025incidents}. The most severe impacts came from two major fires: the Eaton Fire in Altadena \cite{calfire2025eaton} and the Palisades Fire in Pacific Palisades \cite{caramela2025palisades}. Both burned through densely populated wildland–urban interface (WUI) zones, causing significant disruptions and highlighting disparities in disaster response. Based on this setting, we collected 131,268 wildfire-related tweets from California using targeted keywords and location filters from January 1 to 31. After location inference and data cleaning, we retained 28,267 unique tweets across 29 reliably identified cities. These cities include 13 classified as urban and 16 as non-urban, based on population density, metropolitan designation, and socio-infrastructural criteria drawn from ACS classifications and NHGIS\cite{ipums_nhgis_2024} metrics. For mobility data, we applied preprocessing to emphasize consistent movement patterns by filtering out origin–destination (OD) pairs with fewer than five trips. This threshold helped reduce noise from irregular device reporting and focus on habitual behaviors. Daily aggregated trip counts between census tracts were then used for downstream analysis.

\subsection{Model evaluations}
We begin with benchmarking overall model performance across a range of baselines and proposed variants. Table~\ref{tab:overall_performance} presents results in terms of accuracy, recall, and F1-score. Traditional lexicon-based methods such as VADER exhibit poor performance, with an F1-score of only 0.2010, highlighting their limitations in understanding context-rich crisis communication. Transformer-based models, such as RoBERTa and Mistral, offer significant improvements but still underperform in recall, indicating challenges in identifying minority sentiment classes. Our proposed models, particularly those incorporating city-specific adaptation and multimodal fusion, demonstrate consistent gains across all metrics. The \textit{City-Specific-Fusion} model (adaptive model afterwards) achieves the best results, with the highest F1 score (0.5923), recall (0.5652), and accuracy (0.7722), underscoring the benefit of combining local context with behavioral signals.

\begin{table}[htbp]
  \centering
  \caption{Overall performance of sentiment models.}
  \vspace{-10pt}
  \label{tab:overall_performance}
  \begin{tabular}{cccc}
    \toprule
    Model & Accuracy & Recall & F1 \\
    \midrule
    VADER & 0.1954 & 0.3864 & 0.2010 \\
    RoBERTa & 0.6051 & 0.4584 & 0.4444 \\
    Mistral & 0.6792 & 0.5247 & 0.5107 \\
    Global-Pure-Text & 0.7143 & 0.3333 & 0.2778 \\
    City-Specific-Pure-Text & 0.7237 & 0.4180 & 0.3967 \\
    Global-Fusion & \underline{0.7547} & \underline{0.5252} & \underline{0.5549} \\
    City-Specific-Fusion & \textbf{0.7722} & \textbf{0.5652} & \textbf{0.5923} \\
    \bottomrule
  \end{tabular}
\end{table}

Table~\ref{tab:classwise_f1} provides a more detailed breakdown of F1 scores by sentiment class. While all models struggle with positive sentiment, our adaptive model shows the strongest performance across all categories, particularly with a notable improvement in the positive class (Pos\_F1 = 0.4677). This suggests that mobility-enhanced and locality-aware modeling not only boosts average performance but also helps reduce imbalance in sentiment detection. Notably, the high Neu\_F1 (0.8687) indicates that the adaptive model effectively captures neutral expressions, which dominate crisis discourse.

\begin{figure}[htbp]
    \centering
    \includegraphics[width=0.475\textwidth]{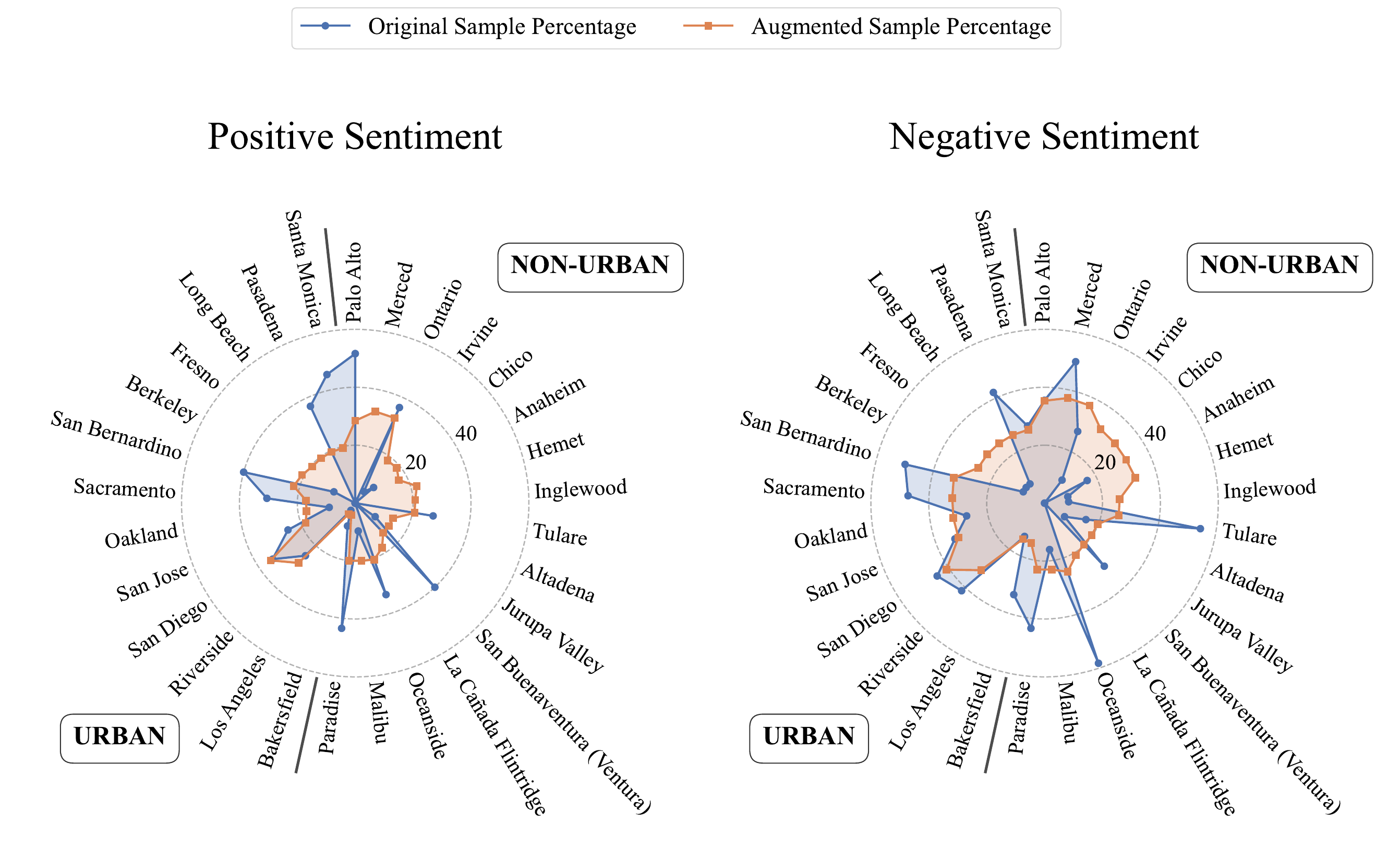}
    \vspace{-20pt}
    \caption{City label distribution comparison.}
    \vspace{-10pt}
    \label{fig:label}
\end{figure}

\begin{table}[htbp]
  \centering
  \caption{Class-wise F1 score breakdown.}
  \vspace{-10pt}
  \label{tab:classwise_f1}
  \begin{tabular}{ccccc}
    \toprule
    Model & F1 & Neg\_F1 & Neu\_F1 & Pos\_F1 \\
    \midrule
    VADER & 0.2010 & 0.2424 & 0.1419 & 0.2186 \\
    RoBERTa & 0.4444 & 0.4485 & 0.7202 & 0.1644 \\
    Mistral & 0.5107 & \textbf{0.5383} & 0.7782 & 0.2155 \\
    Global-Fusion & \underline{0.5549} & \underline{0.4790} & \underline{0.8524} & \underline{0.3333} \\
    City-Specific-Fusion & \textbf{0.5923} & 0.4404 & \textbf{0.8687} & \textbf{0.4677} \\
    \bottomrule
  \end{tabular}
  \vspace{-10pt}
\end{table}
To further explore model fairness, we analyze the distribution of sentiment labels across urban and non-urban city groups. As shown in Figure~\ref{fig:label}, we visualize the proportions of positive and negative sentiment before and after applying data augmentation. In the original sample distribution (blue), significant class imbalance and high variance are observed, especially in non-urban cities. For example, cities like La Cañada Flintridge and Chico exhibit no negative-sentiment tweets (both have 0\% negative sentiment). In contrast, the augmented distribution (orange) provides more balanced sentiment coverage across all cities. This effect is evident in the negative sentiment class, where many non-urban locations gain clearer representation post-augmentation.
This shift likely reflects their proximity to wildfire zones and the heightened emotional impact of such exposure. By enriching data in under-sampled areas, our augmentation strategy narrows the gap between urban and non-urban groups. For instance, while urban centers such as Riverside (before: positive: 25\%, negative: 41.67\%  after: positive: 28.45\%, negative: 31.9\%) and San Diego (before: positive: 34.48\%, negative: 44.83\% after:positive: 35.24\%, negative: 40.95\% ) already had relatively balanced sentiment distributions, non-urban cities such as Tulare (before: positive: 27.27\%, negative: 54.55\% after: positive: 20.78\%, negative: 25.97\%) 
and Merced (before: positive: 0\%, negative: 50\% after: positive:,34\% negative:39\%) show reduced gaps between polarity classes following augmentation. 

\subsection{Sentiment mapping}
\begin{figure*}[htbp]
    \centering
    \includegraphics[width=0.82\textwidth]
    {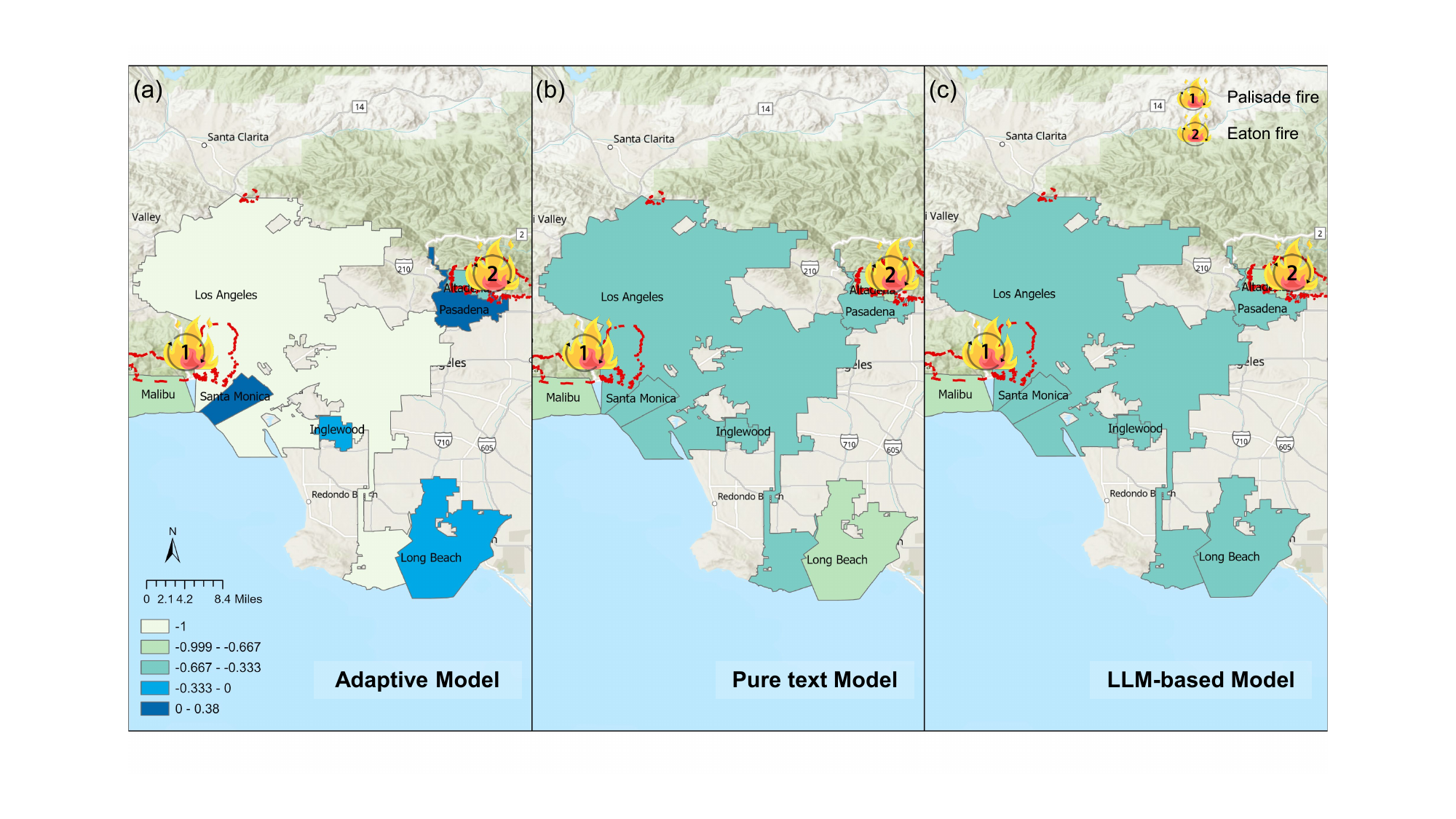}
    \vspace{-5pt}
    \caption{Estimated sentiment distribution among (a) adaptive model, (b) pure text model, and (c) LLM-based model.}
    \vspace{-5pt}
    \label{fig: sentiment}
\end{figure*}

To assess public sentiment trends over time and space, we introduce the metric of \textit{accumulative sentiment}, which smooths temporal fluctuations and highlights overall emotional responses to wildfire events. Formally, the accumulative sentiment score at time $t$ is defined as:
\begin{equation}
\text{Acc\_sentiment}(t) = 
\frac{\sum_{i=1}^{t} (\text{pos}_i - \text{neg}_i)}{\sum_{i=1}^{t} (\text{pos}_i + \text{neg}_i)}
\end{equation}
where $\text{pos}_i$ and $\text{neg}_i$ represent the number of positive and negative tweets, respectively. This metric captures the evolving emotional tone over time, smoothing short-term volatility from low tweet volumes \cite{li2020leveraging}. Using this metric, we visualized city-level sentiment distributions produced by three models: our proposed adaptive model, a fine-tuned text-only transformer (Global-Pure-Text) model, and a large language model based on Mistral (Figure \ref{fig: sentiment}).

The adaptive model produces more geographically diverse sentiment estimates. In the broader Los Angeles city, sentiment drops to $-1$ under the adaptive model, compared to $-0.5$ (Global-Pure-Text) and $-0.33$ (LLM). In coastal areas such as Malibu, sentiment estimates range from $-0.58$ (LLM) to $-0.70$ (adaptive). For cities near the Eaton fire, sentiment improves under the adaptive model, with Pasadena shifting from $-0.61$ (LLM) and $-0.47$ (Global-Pure-Text) to $+0.25$ (adaptive), and Altadena from $-0.65$ (Global-Pure-Tex and LLM) to $-0.40$ (adaptive). Urban–suburban sentiment gradients are also more pronounced. For example, Santa Monica maintains a relative neutral tone ($0.21$), while adjacent Malibu shows more negativity ($-0.70$) in the adaptive model. Pasadena and Altadena show differing sentiment directions despite their proximity. These distinctions are less visible in the other models, demonstrating the higher spatial sensitivity of the adaptive approach.

\subsection{Word-to-action analysis}
\begin{figure}[htbp]
    \centering
    \includegraphics[width=0.475\textwidth]{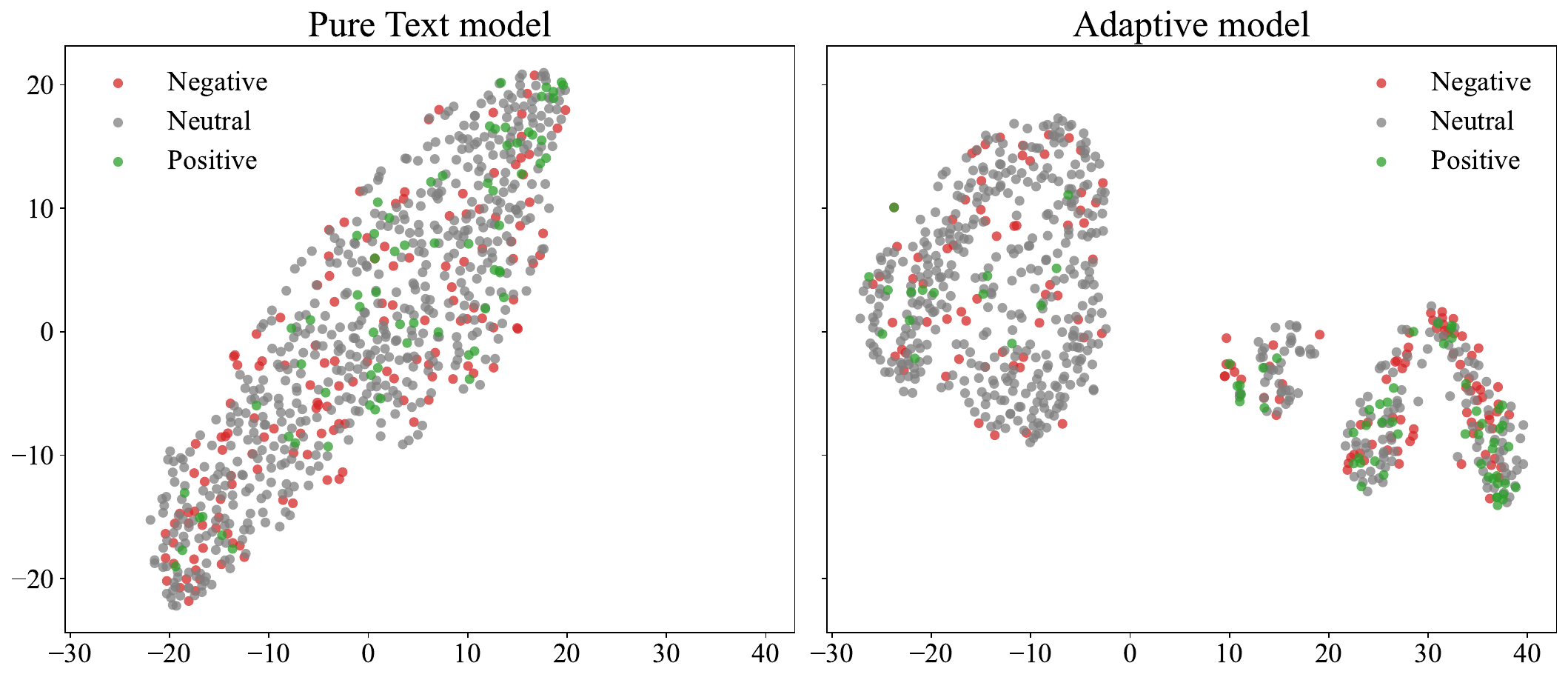}
    \vspace{-10pt}
    \caption{Effect of mobility embedding on sentiment representation via t-SNE.}
    \vspace{-20pt}
    \label{fig:tsne}
\end{figure}

Understanding the interplay between public sentiment and physical mobility is critical during disasters, as it reveals how emotional responses influence real-world behavior \cite{xu2025impact, li2020leveraging}. To explore this relationship, we integrate both sentiment and mobility signals into our framework and examine their combined influence. We begin with a t-SNE--based visualization to evaluate how the inclusion of mobility embeddings affects sentiment representation, comparing the adaptive model (with mobility) against the Global-Pure-Text model (without mobility).
As shown in Figure~\ref{fig:tsne}, adaptive model produce more distinct clusters for negative, neutral, and positive tweets. In contrast, models without mobility information yield entangled and compact sentiment representations, suggesting limited semantic separation between classes. 

\begin{figure*}[htbp]
    \centering
    \includegraphics[width=\textwidth]{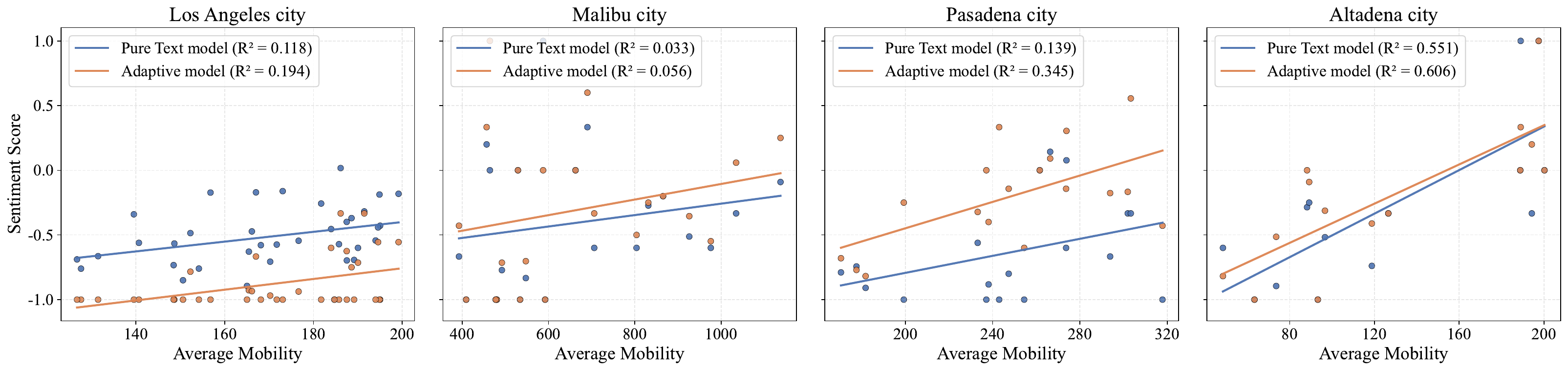}
    \vspace{-20pt}
    \caption{Correlation between sentiment score and average mobility across selected cities.}
    \vspace{-10pt}
    \label{fig:regression}
\end{figure*}

To further quantify this relationship, we conduct a correlation analysis across four representative cities: Los Angeles, Malibu, Pasadena, and Altadena, capturing one urban and one non-urban pair for each wildfire, as shown in Figure~\ref{fig:regression}. In each subplot, the orange dots and regression lines represent correlations derived from our adaptive model, while the blue ones correspond to the Global-Pure-Text model. Overall, we observe consistently positive correlations between average sentiment and mobility, indicating that periods of more positive online expression tend to align with increased physical movement. The adaptive model shows stronger and more stable correlations than the Global-Pure-Text model, most notably in suburban Altadena ($R^2 = 0.606$ vs.~0.551). Although the improvements in urban Los Angeles are relatively modest, the general pattern suggests that incorporating mobility signals enables the model to better capture co-occurring behavioral shifts, providing a more contextually grounded understanding of community dynamics during crises.

\subsection{Ablation studies}
To evaluate the impact of our architectural and training design choices, we conduct a set of ablation studies focusing on city-level representation, training strategies, and multimodal input handling. 
\begin{figure}[htbp]
    \centering
    \includegraphics[width=0.475\textwidth]
    {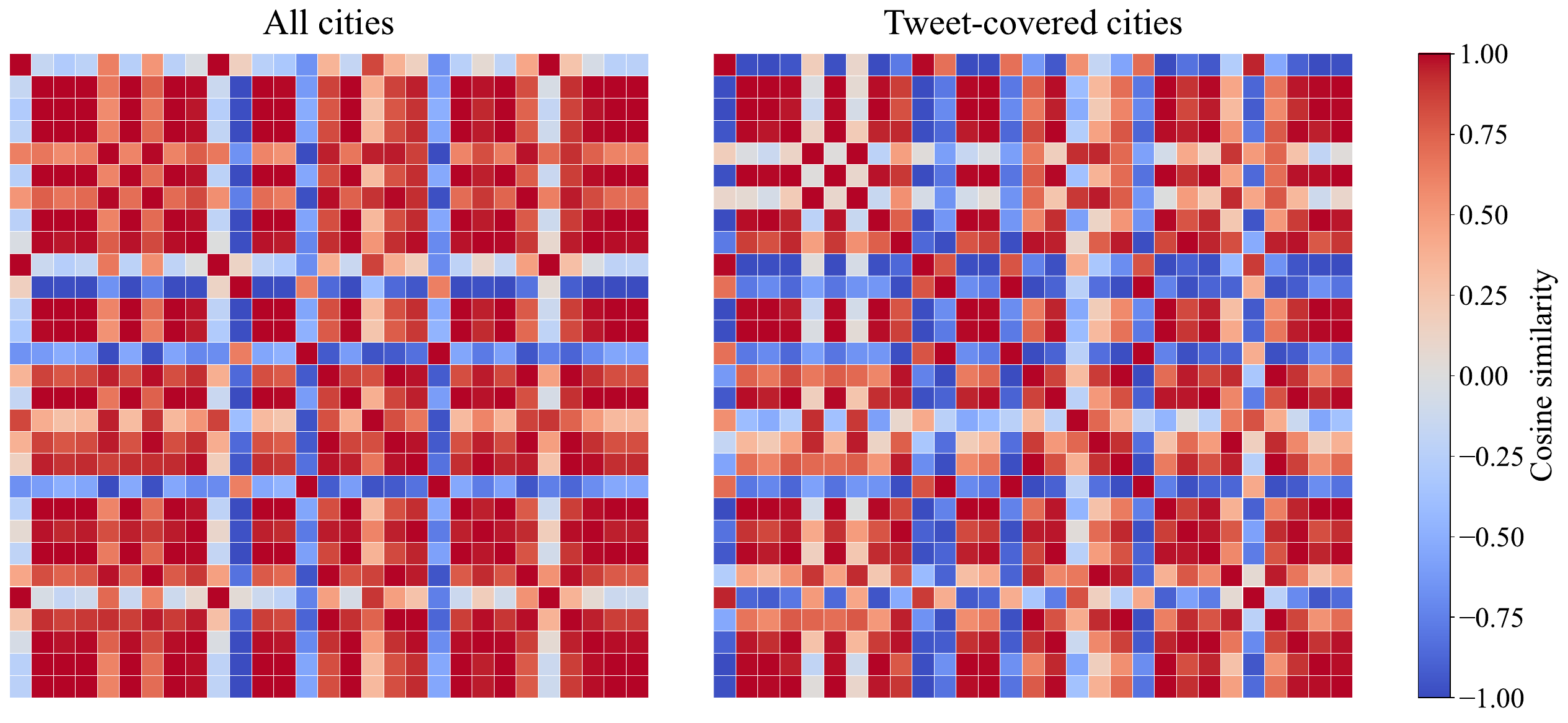}
    \vspace{-20pt}
    \caption{City-wide learning comparison.}
    \vspace{-10pt}
    \label{fig:A2SL}
\end{figure}

\noindent\textbf{\textit{City-wide representation generalization.}} 
We first examine the role of broader city representation learning in shaping stable city embeddings. Figure~\ref{fig:A2SL} visualizes the cosine similarity matrices for city representations under two training configurations: one using all 938 California cities from the socioeconomic dataset (left), and the other restricted to the 29 tweet-covered cities(right). When training is performed across the full set of cities, the similarity structure exhibits smoother gradients and coherent cluster blocks, with higher intra-cluster similarity and clearer separation between distinct city groups. In contrast, the tweet-only training condition yields a noisier and less interpretable matrix, characterized by more erratic red–blue alternations and fragmented similarity zones. These results indicate that including all cities not only regularizes representation learning but also enables the model to align city semantics over a larger spatial context. 

\begin{figure}[htbp]
    \centering
    \includegraphics[width=0.375\textwidth]{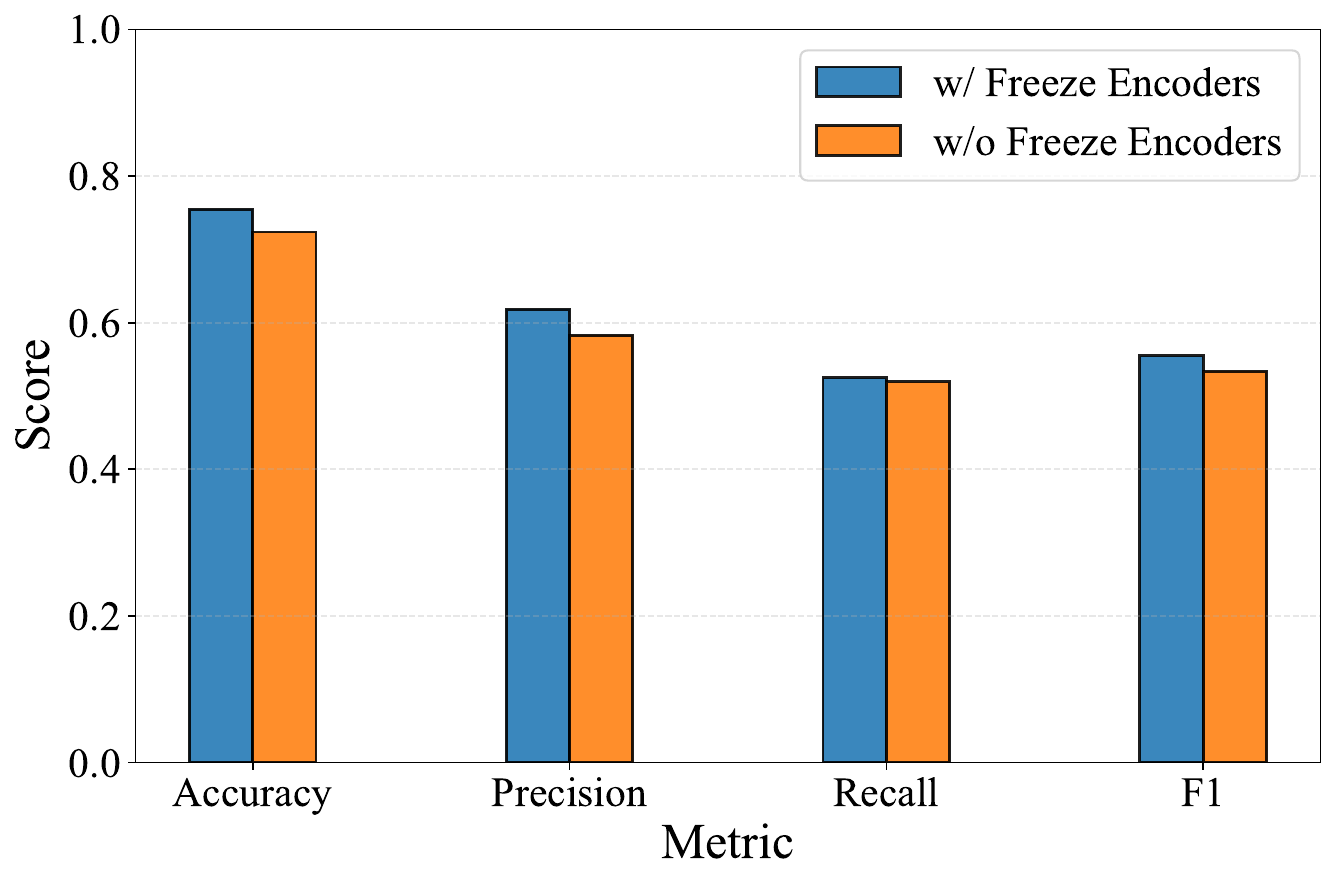}
    \vspace{-10pt}
    \caption{Training strategy comparison.}
    \vspace{-15pt}
    \label{fig:freeze}
\end{figure}

\noindent\textbf{\textit{Training strategy for model stability.}} 
We then compare two supervised refinement strategies to assess how different levels of encoder flexibility influence learning. Based onFigure~\ref{fig:freeze}, freezing both the text and mobility encoders while training only the perception layer consistently yields better performance across all metrics. 
This finding reinforces that, under limited training data, fine-tuning only the final layer preserves general representations while reducing the risk of overfitting. Conversely, fully unfreezing the encoders leads to diminished scores, suggesting that excessive parameter updates introduce instability during the adaptation phase. 

\noindent\textbf{\textit{Effectiveness of multimodal fusion.}} 
Finally, Table~\ref{tab:Ablation} presents a comparison of two input designs for incorporating mobility signals: a prompt-based approach that verbalizes mobility as textual input, and a multimodal fusion strategy that treats mobility as a separate numerical modality. Across both global and city-specific settings, fusion-based models consistently outperform their prompt-based counterparts, with F1-scores improving from $0.35$ to $0.56$ in the city-specific configuration. These results underscore that treating mobility as structured input rather than auxiliary text allows the model to capture richer behavioral signals and integrate them more effectively into sentiment inference.

\section{Discussion}
\subsection{Fairness-aware representation improvement}

Our proposed adaptive cross-city learning framework demonstrates notable improvements in both predictive performance and representational fairness. By integrating city similarity-based data augmentation with mobility-aware sentiment modeling, the framework effectively transfers knowledge from data-rich urban centers to less-represented regions. This augmentation process leverages structural and socioeconomic similarities between cities to expand the available training data, ensuring a more robust representation for undersampled locations and improving model generalizability~\cite{luo2025great}.

Crucially, this strategy help reduce participation biases and enable the framework to capture more geographically diverse and demographically relevant sentiment signals. Compared to traditional models, our model reveals sharper emotional distinctions between adjacent communities, particularly those affected by overlapping wildfire hazards. For example, in Los Angeles city, where the fire zones of the Eaton and Palisades wildfires intersect, our model estimates the lowest sentiment score ($-1$). This potentially corresponds to heightened evacuation complexity and cascading disruptions experienced by residents in this high-risk zone, as the City of Los Angeles declared a state of emergency on January~7 \cite{hancock2025california} and maintained evacuation orders until January~27 \cite{lloyd2025palisades}, reflecting prolonged community stress. Similarly, coastal cities such as Pacific Palisades ($-1$) and Malibu ($-0.7$) exhibit markedly lower sentiment estimates, consistent with the extensive property losses and concentrated public attention these areas experienced.

\subsection{Equitable disaster response enhancement}

Beyond algorithmic performance, the model’s fine-grained outputs offer actionable insights for emergency management and equitable resource allocation. For example, in Malibu, where one-third of the city was reported lost and homes along the Pacific Coast Highway and in the Big Rock neighborhood were destroyed \cite{bbc2025lafires}, the adaptive model estimates heightened negative sentiment, indicating limited evacuation routes and delayed emergency notifications.

In the case of the Eaton Fire, our model highlights divergent emotional responses across neighboring communities. While Altadena remains relatively negative ($-0.45$), Pasadena shows a more positive sentiment ($0.25$), likely reflecting stronger emergency response capacity and coordinated community support, such as volunteer assistance efforts at Santa Anita Park \cite{rivera2025eaton}. As of July~22,~2025, the fire had caused 19 deaths and left 22 individuals missing \cite{pasadenanow2025eaton, pasadenastar2025eaton}. Most victims resided west of Lake Avenue in Altadena, a predominantly Black neighborhood that reportedly received evacuation orders several hours later than areas east of Lake Avenue \cite{beckett2025lafires}. This delay possibly contributes to the heightened negative sentiment expressed by residents in the area.

These observations are further supported by our word-to-action analysis, which reveals stronger correlations between sentiment and mobility in suburban regions. This suggests that emotional responses are more likely to manifest as real-world behavioral changes in these communities. Such patterns offer valuable guidance for identifying areas requiring faster intervention, enhanced communication infrastructure, or targeted support strategies. Among the four cities studied, Altadena demonstrates the highest correlation between sentiment and mobility, indicating that public sentiment more directly influences behavioral outcomes. This implies that decision-makers may prioritize psychological support and communication interventions to strengthen community resilience in similar contexts. In contrast, Los Angeles exhibits weaker correlations, suggesting more diffuse or heterogeneous behavioral responses. As such, greater effort may be required to ensure effective public safety engagement in densely populated urban centers.

\begin{table}[t]
  \centering
  \caption{Ablation study of different inputs.}
  \vspace{-10pt}
  \label{tab:Ablation}
  \begin{tabular}{cccc}
    \toprule
    Model & Accuracy & Recall & F1 \\
    \midrule
    Global-Prompt & 0.7224 & 0.3575 & 0.3281 \\
    City-Specific-Prompt & 0.7412 & 0.4097 & 0.4184 \\
    Global-Fusion & \underline{0.7547} & \underline{0.5252} & \underline{0.5549} \\
    City-Specific-Fusion & \textbf{0.7722} & \textbf{0.5652} & \textbf{0.5923} \\
    \bottomrule
  \end{tabular}
  \vspace{-0.5cm}
\end{table}

\section{Conclusion}

This paper presents an adaptive cross-city fusion framework designed to enhance representational equity and interpretability in disaster sentiment analysis. By integrating textual data with mobility-informed behavioral features and city similarity-based data augmentation, the model captures localized emotional dynamics that traditional models often overlook. Our case study demonstrates the framework’s ability to uncover fine-grained sentiment variations across diverse urban and non-urban geographies. 
The model also identifies sharper emotional boundaries between adjacent areas such as Santa Monica and Malibu, revealing gaps in perceived safety, digital infrastructure, and institutional responsiveness. Additionally, our proposed model exhibits a stronger positive correlation between digital sentiment and real-world behavioral patterns, with notable variations across cities. These insights enable more informed, context-sensitive strategies for equitable disaster preparedness and response.

\bibliographystyle{ACM-Reference-Format}
\bibliography{main}

\section{Acknowledgment}
The authors gratefully acknowledge Dr. Lingyao Li (University of South Florida) and Dr. Yang Fan (University of South Carolina) for collecting the data and providing access to the dataset through the Brandwatch platform, which was essential to the analyses presented in this study.

\clearpage

\section{Appendices}
\label{sec:appendix}
\subsection*{Limitations}
Although our proposed framework demonstrates promising advancements in representational fairness and sentiment interpretability, several limitations warrant further discussion.

First, the quality and representativeness of geo-tagged social media data introduce fundamental challenges. Our data collection strategy relies on keyword filtering, which, while efficient, may exclude relevant tweets that use nonstandard, community-specific, or evolving crisis language. Additionally, despite preprocessing and noise-reduction efforts, location metadata is often sparse or imprecise, limiting our ability to ground sentiment predictions in fine-grained geographies. While the city similarity-based data augmentation strategy helps mitigate data sparsity, it also introduces assumptions about structural and socioeconomic comparability that may not fully capture local linguistic or cultural variations.

Second, sentiment on social media platforms is inherently ambiguous. Our model is designed to classify emotional polarity, but crisis-related tweets often blend personal experience, factual reporting, and institutional messaging. As a result, some sentiment scores may reflect narrative tone rather than genuine emotional states. This ambiguity poses challenges for interpretation, especially in high-stakes environments where distinguishing between concern, criticism, and information is crucial.

Third, while the inclusion of mobility statistics adds valuable behavioral context, these signals are aggregated and do not capture real-time, individual-level movement dynamics. Moreover, our modeling assumes that mobility fluctuations are predominantly driven by wildfire exposure, yet other latent factors, such as seasonal variation, public health advisories, or economic pressures, may confound these patterns. Disentangling disaster-specific mobility from unrelated behavioral noise remains a challenge for future refinement.

Fourth, the framework operates at the city level, which constrains spatial granularity and may obscure sentiment variation within cities. Neighborhood-level disparities—especially in larger metropolitan areas—are likely underrepresented. Addressing this would require access to more detailed geographic annotations and disaggregated behavioral data, which are not uniformly available.

Fifth, while our multimodal perception layer integrates text and mobility features effectively, the fusion mechanism is relatively static. Future work could explore more dynamic attention-based methods to model intermodal dependencies more adaptively. Additionally, although our City-wide Learning model can quantify inter-city similarity, it sometimes produces extreme similarity values for specific city pairs. This limitation stems in part from our sampling strategy, which is currently driven by wildfire exposure alone. Incorporating additional features—such as transit infrastructure or population mobility baselines—could support more robust and nuanced similarity estimation.

Finally, the generalizability of our findings beyond the January 2025 Southern California wildfires remains an open question. Different disaster types may have distinct emotional and behavioral responses, requiring the adaptation of our modeling assumptions and input modalities. Extending this framework to broader spatiotemporal contexts and integrating richer modalities, such as imagery or social network structure, may further enhance both its applicability and resilience. 

\subsection*{Ethics}
This study utilizes social media data obtained under an academic research license, subject to the platform’s data use agreement and ethical research guidelines. While the data are not publicly available, access was granted for non-commercial, scholarly purposes and complied with all applicable terms of service and institutional data governance policies. All records were anonymized and processed in aggregate to protect user privacy, and no personally identifiable information (PII) was collected, stored, or disclosed. Human annotations for sentiment labeling were performed by trained researchers with informed consent. The study adheres to ethical standards of digital research, emphasizing fairness, transparency, and responsible interpretation to avoid reinforcing social or spatial inequities in disaster-related analyses and decision-making.

\subsection*{Sentiment prompt}
\begin{lstlisting}[language=Python]
SENTIMENT_PROMPT = """
Task:
You are a careful annotator for social-media posts. Your job is to assign one sentiment label to each tweet: negative, neutral, or positive. Think through the decision silently. Do not reveal your reasoning. Output only the final JSON object.

Input:Tweet Text: {tweet}

Instructions:
Annotation policy (apply in this order):
1. Negative if the tweet conveys complaint, fear, loss, harm, danger, evacuation stress, property/health damage, anger, blame, or explicitly negative emotion -- even if mixed with facts.
2. Positive if it clearly expresses relief, gratitude, appreciation, praise, successful containment/help, hopeful outlook or other explicit positive emotion.
3. Neutral if it's informational/ambiguous (e.g., resource provided such as a website/shelter link, questions without clear emotion, logistics, "only hashtags/keywords with no substantive content," or irony that cannot be resolved from context).
4. Emojis, intensifiers, negations, and sarcasm matter only when unambiguous from the text.
5. Hashtags and mentions count only if they clearly express sentiment (e.g., #grateful).
6. When unsure, choose neutral.


Few-shot examples:
<Positive tweet examples> 

<Negative tweet examples> 

<Neutral tweet examples> 

Output:
Output must be in strict JSON format with the following structure:
{{
    "sentiment": "<negative|neutral|positive>"
    "confidence": 0-1
}}

"""
\end{lstlisting}

\subsection*{Descriptive statistics of the variables.}

Guided by prior research \cite{ma2025analyzing} and the Social Vulnerability Index (SVI) framework—which spans socioeconomic status, household composition, racial and ethnic minority status, and housing and transportation—we first designated 22 tract-level ACS indicators as core controls, given their established importance for measuring social vulnerability and disaster risk. Building on this foundation, we assembled an extended pool of candidate predictors for the city-wide learning layer and applied a Variance Inflation Factor (VIF) screening procedure to the remaining 79 non-core variables (Table~\ref{tab:vif_final_features}). 

\begin{table}[htbp]
\centering
\small
\caption{Final VIF values for the selected ACS tract-level features.}
\label{tab:vif_final_features}
\begin{tabular}{lr}
\toprule
Feature & Final VIF \\
\midrule
pct\_occupation\_healthcare\_support & 4.8787 \\
pct\_commute\_carpool               & 4.5577 \\
pct\_occupation\_business           & 3.9466 \\
pct\_married\_now                   & 3.7147 \\
pct\_rent\_1000\_1499               & 3.7038 \\
pct\_rent\_1500\_1999               & 3.6981 \\
pct\_complete\_kitchen              & 3.5558 \\
pct\_occupation\_computer           & 3.5024 \\
pct\_occupation\_social\_service    & 3.4445 \\
pct\_rent\_2000\_more               & 3.2875 \\
pct\_occupation\_protective         & 3.1240 \\
pct\_industry\_construction         & 3.0623 \\
pct\_graduate\_degree               & 3.0438 \\
pct\_occupation\_education          & 3.0158 \\
pct\_households\_living\_alone      & 2.8306 \\
pct\_occupation\_food\_service      & 2.8292 \\
pct\_rent\_500\_999                 & 2.7753 \\
occupants\_per\_room                & 2.7522 \\
pct\_units\_2                       & 2.6711 \\
pct\_units\_1\_attached             & 2.6203 \\
pct\_units\_1\_detached             & 2.5293 \\
pct\_occupation\_architecture       & 2.2026 \\
pct\_occupied\_housing              & 2.1364 \\
pct\_occupation\_legal              & 1.9880 \\
pct\_commute\_public\_transit       & 1.9229 \\
median\_rooms                       & 1.8662 \\
pct\_occupation\_management         & 1.7596 \\
pct\_commute\_walk                  & 1.6206 \\
pct\_rent\_under\_500               & 1.5962 \\
pct\_complete\_plumbing             & 1.5530 \\
pct\_no\_telephone                  & 1.5415 \\
households\_cohabiting              & 1.5331 \\
pct\_units\_3\_4                    & 1.4679 \\
pct\_native\_alone                  & 1.3940 \\
pct\_heating\_fuel\_none            & 1.3876 \\
pct\_heating\_fuel\_oil             & 1.1258 \\
pct\_heating\_fuel\_wood            & 1.0171 \\
\bottomrule
\end{tabular}
\end{table}

We iteratively removed the variable with the highest VIF until all surviving predictors had VIF values below the conventional threshold of 5, thereby mitigating multicollinearity and improving model stability. This process yielded 37 additional predictors, which we combined with the 22 core ACS indicators to obtain a final set of 59 tract-level variables for the city-wide learning layer. These variables serve as structured inputs for generating city embeddings in our adaptive model, enabling it to represent social and infrastructural heterogeneity across both urban and non-urban areas. Descriptive statistics for the tract-level variables are summarized in Table~\ref{tab:ca_tract_demographics_stats}.

\begin{table*}[htbp]
\centering
\small
\caption{Summary statistics of census-tract-level demographic features in California.}
\begin{tabular}{lrrrr lrrrr}
\toprule
Feature & Mean & Std & Min & Max & Feature & Mean & Std & Min & Max \\
\midrule
urban\_rate & 92.37 & 23.78 & 0.0 & 100.0 & pct\_occupation\_education & 5.71 & 4.12 & 0.0 & 44.4 \\
median\_household\_income & 99732.02 & 46278.62 & 9417.0 & 250001.0 & pct\_occupation\_healthcare\_support & 9.83 & 5.50 & 0.0 & 71.4 \\
mean\_household\_income & 127680.52 & 64137.98 & 13384.0 & 603656.0 & pct\_occupation\_protective & 4.99 & 3.55 & 0.0 & 100.0 \\
pct\_black\_alone & 5.60 & 8.50 & 0.0 & 90.6 & pct\_occupation\_food\_service & 4.71 & 4.12 & 0.0 & 56.2 \\
pct\_hispanic\_latino & 25.09 & 18.87 & 0.0 & 100.0 & pct\_commute\_carpool & 9.52 & 6.08 & 0.0 & 100.0 \\
pct\_white\_not\_hispanic & 6.64 & 7.04 & 0.0 & 78.3 & pct\_commute\_public\_transit & 3.65 & 6.15 & 0.0 & 78.0 \\
pct\_asian\_alone & 14.41 & 16.28 & 0.0 & 100.0 & pct\_commute\_walk & 2.57 & 5.35 & 0.0 & 100.0 \\
fem\_pct & 49.87 & 5.30 & 0.0 & 100.0 & pct\_complete\_plumbing & 1.00 & 1.56 & 0.0 & 20.5 \\
male\_pct & 50.13 & 5.30 & 0.0 & 100.0 & occupants\_per\_room & 3.36 & 4.68 & 0.0 & 78.9 \\
pct\_hs\_grad\_higher & 83.85 & 14.00 & 0.0 & 100.0 & pct\_no\_telephone & 1.13 & 2.39 & 0.0 & 48.0 \\
less\_9th\_pct & 9.09 & 9.53 & 0.0 & 78.2 & pct\_units\_1\_attached & 4.70 & 6.55 & 0.0 & 69.1 \\
pct\_bachelor\_higher & 35.30 & 21.79 & 0.0 & 100.0 & pct\_units\_3\_4 & 3.68 & 8.68 & 0.0 & 100.0 \\
unemployment\_rate & 6.58 & 4.21 & 0.0 & 66.7 & median\_rooms & 203.43 & 230.57 & 0.0 & 1956.0 \\
poverty\_rate & 12.57 & 9.73 & 0.0 & 100.0 & pct\_occupied\_housing & 7.31 & 8.98 & 0.0 & 100.0 \\
eld\_65p\_pct & 15.71 & 8.88 & 0.0 & 100.0 & pct\_industry\_construction & 4.50 & 4.02 & 0.0 & 51.6 \\
youth\_18l\_pct & 21.50 & 7.40 & 0.0 & 72.7 & pct\_rent\_under\_500 & 3.35 & 6.95 & 0.0 & 95.8 \\
avg\_household\_size & 2.96 & 0.69 & 1.04 & 7.19 & pct\_rent\_500\_999 & 9.56 & 13.39 & 0.0 & 100.0 \\
pct\_owner\_occupied & 55.77 & 24.77 & 0.0 & 100.0 & pct\_rent\_1000\_1499 & 19.27 & 17.23 & 0.0 & 100.0 \\
pct\_foreign\_born & 0.33 & 1.44 & 0.0 & 71.5 & pct\_rent\_1500\_1999 & 21.70 & 16.14 & 0.0 & 100.0 \\
per\_capita\_income & 46813.76 & 27489.78 & 482.0 & 278590.0 & pct\_rent\_2000\_more & 17.83 & 14.81 & 0.0 & 100.0 \\
median\_age & 38.76 & 7.96 & 10.8 & 82.3 & pct\_heating\_fuel\_oil & 0.26 & 1.73 & 0.0 & 42.9 \\
total\_population & 4311.11 & 1739.09 & 0.0 & 38907.0 & households\_cohabiting & 18.61 & 26.91 & 0.0 & 342.0 \\
built\_2014\_pct & 0.37 & 1.34 & 0.0 & 35.5 & pct\_households\_living\_alone & 4.67 & 4.47 & 0.0 & 100.0 \\
pct\_native\_alone & 1.03 & 1.98 & 0.0 & 73.2 & pct\_married\_now & 2.69 & 1.94 & 0.0 & 17.1 \\
pct\_occupation\_management & 2.51 & 6.61 & 0.0 & 72.1 & pct\_graduate\_degree & 4.79 & 4.28 & 0.0 & 100.0 \\
pct\_occupation\_business & 6.82 & 4.78 & 0.0 & 100.0 & pct\_units\_1\_detached & 5.59 & 7.16 & 0.0 & 82.9 \\
pct\_occupation\_computer & 8.58 & 5.11 & 0.0 & 42.3 & pct\_units\_2 & 11.99 & 17.83 & 0.0 & 98.6 \\
pct\_occupation\_architecture & 2.59 & 2.41 & 0.0 & 55.6 & pct\_complete\_kitchen & 5.60 & 5.89 & 0.0 & 46.7 \\
pct\_occupation\_social\_service & 5.85 & 4.26 & 0.0 & 31.8 & pct\_heating\_fuel\_wood & 0.01 & 0.16 & 0.0 & 7.1 \\
pct\_occupation\_legal & 2.86 & 3.68 & 0.0 & 41.2 & pct\_heating\_fuel\_none & 0.92 & 1.72 & 0.0 & 27.2 \\
wfir\_risk\_score & 2.23 & 1.68 & 0 & 5 &  &  &  &  &  \\
\bottomrule
\end{tabular}
\label{tab:ca_tract_demographics_stats}
\end{table*}

\subsection*{Annotation and model selection}

Tweets were collected from Brandwatch using the following criteria:

\begin{itemize}
    \item Search terms: \textit{wildfire}, \textit{bushfire}, \textit{California fire}, \textit{CAfire}, \textit{CALfire}
    \item Search location: California
    \item Search period: January 1 2025 -- January 31  2025
\end{itemize}

To establish a high-quality ground truth for evaluating language models, we first conducted a structured manual annotation process. This enabled a rigorous comparison of large language models (LLMs) for sentiment prediction on wildfire-related tweets.

\textbf{Annotation framework.} Two trained annotators labeled 2,000 tweets to classify sentiment as \textit{positive} (1), \textit{neutral} (0), or \textit{negative} (-1). Negative sentiment primarily reflected emotional distress such as fear, grief, or anger, which was often linked to property loss, evacuation, or emergency delays. Positive sentiment captured support, gratitude, or community solidarity. Neutral sentiment was assigned to tweets lacking affective tone or conveying factual information. To evaluate annotation consistency, we measured inter-annotator reliability using \textbf{Krippendorff’s Alpha} ($\alpha$), a robust coefficient designed for nominal data and multiple coders:

\begin{equation} 
\alpha = 1 - \frac{D_o}{D_e} \label{eq:kripp_alpha} \end{equation}
where $D_o$ represents the \textit{observed disagreement} among annotators and 
$D_e$ represents the \textit{expected disagreement} under conditions of random annotation. The observed disagreement is computed as 
$D_o = \frac{1}{N} \sum_{i=1}^{N} \delta(a_{i1}, a_{i2})$, 
where $N$ is the total number of annotated items and $a_{i1}$ and $a_{i2}$ denote the sentiment labels assigned by the two annotators for item $i$. The function $\delta(a,b)=1$ when the annotations differ and $0$ when they agree. The expected disagreement is defined as 
$D_e = \sum_{c_1 \ne c_2} p(c_1)\, p(c_2)$,
where $p(c)$ is the proportion of annotations assigned to category $c$.

A higher value of $\alpha$ indicates stronger agreement after adjusting for chance, while values near zero suggest agreement consistent with random labeling. The final reliability score (0.763) reflects substantial consistency and interpretive alignment among annotators, ensuring that the labeled dataset provides a reliable foundation for model evaluation.

\textbf{LLM evaluation.} We evaluated a suite of eight LLMs, including open-source models deployable on local GPUs (e.g., \texttt{Mistral-8B}, \texttt{DeepSeek-8B}, \texttt{Llama-8B}) as well as proprietary API-based systems (e.g., \texttt{GPT-4o},\texttt{GPT-5}, \texttt{Gemini~2.5}). Each model was assessed using a standardized three-step prompting workflow to ensure consistency across evaluations:

\begin{itemize}
    \item Location detection -- further identifying tweets originating from or referencing the study region.
    \item Event relevance identification --further filtering messages to retain only wildfire-related content.
    \item Sentiment classification -- assigning each tweet to one of three sentiment categories.
\end{itemize}

All models were evaluated under both zero-shot and few-shot configurations. To preserve the authenticity and inherent noise of the social media data, only minimal preprocessing was applied to the collected tweets.

\begin{table}[h!]
\centering
\caption{Model performance summary}
\label{tab:model_summary}
\begin{tabular}{lcc}
\toprule
\textbf{Model} & \textbf{Accuracy} & \textbf{F1} \\
\midrule
Mistral-8B-Instruct-2410        & 75.4\% &0.76 \\ 
DeepSeek-R1-Distill-Llama-8B    & 74.7\% & 0.75 \\ 
Gemini 2.5                      & 65.0\% & 0.469 \\
GPT-5 Instant                   & 60.0\% & 0.600 \\ 
GPT-4o-mini                     & 60.0\% & 0.413 \\
GPT-4o                          & 55.0\% & 0.393 \\
GPT-5 Thinking                  & 55.0\% & 0.413 \\
LLaVA-v1.6-Vicuna-7B            & 55.5\% & 0.53 \\ 
\bottomrule
\end{tabular}
\end{table}

Based on Table \ref{tab:model_summary}, although multimodal models such as \texttt{LLaVA} demonstrated versatility, they performed less effectively on this text-only dataset. Among all evaluated systems, \texttt{Mistral-8B} achieved the highest and most stable accuracy, closely followed by \texttt{DeepSeek}. We therefore selected \texttt{Mistral-8B} as the baseline sentiment classifier for downstream analysis due to its strong performance, efficient inference on A100 GPUs, and architecture that is well suited for fine-tuning under constrained computational settings. 

Importantly, our evaluation also reveals a broader insight: while tested LLMs achieve competitive accuracy, they nonetheless exhibit systematic biases in sentiment interpretation—particularly when processing emotionally nuanced, community-specific, or geographically contextualized language. These biases underscore the need for fairness-aware modeling approaches such as ours, which explicitly incorporate demographic and spatial diversity into disaster communication analysis.

\end{document}